\documentclass{amsart}

\usepackage{graphicx}
\usepackage{geometry}
\usepackage{amssymb}
\usepackage{amsmath}
\usepackage{amsfonts}
\usepackage[obeyFinal,textsize=footnotesize]{todonotes}
\usepackage{array}
\usepackage{multirow}
\usepackage{makecell}
\usepackage{arydshln}
\usepackage{multirow,bigdelim}
\usepackage{mathtools}
\usepackage{verbatim}
\usepackage{tabu}
\usepackage[font=small]{caption}
\usepackage{url}
\usepackage[hidelinks]{hyperref}
\usepackage{hhline}
\usepackage{fancyvrb}
\usepackage{comment}
\usepackage{hhline}
\usepackage{algorithm,algpseudocode}
\usepackage{setspace}
\usepackage{enumitem}

\newtheorem{theorem}{Theorem}[section]
\newtheorem{proposition}[theorem]{Proposition}
\newtheorem{corollary}{Corollary}[theorem]
\newtheorem{lemma}[theorem]{Lemma}
\theoremstyle{definition}
\newtheorem{definition}[theorem]{Definition}

\theoremstyle{remark}

\numberwithin{equation}{section}
\newcommand{\RN}[1]{%
  \textup{\uppercase\expandafter{\romannumeral#1}}%
}
\newcommand{\rn}[1]{%
  {\lowercase\expandafter{(\romannumeral#1)}}%
}

\newcommand{\plv}{Painlev\'{e} }

\makeatletter
\renewcommand*\env@matrix[1][*\c@MaxMatrixCols c]{%
  \hskip -\arraycolsep
  \let\@ifnextchar\new@ifnextchar
  \array{#1}}
\makeatother

\begin{document}

\title[The smallest eigenvalue of the L$\beta$E and J$\beta$E]{The smallest eigenvalue of $\beta$-Laguerre and $\beta$-Jacobi ensembles and multivariate orthogonal polynomials}

\author{Sungwoo Jeong}

\address{Department of Mathematics, Cornell University, Ithaca, New York}
\email{sjeong@cornell.edu}

\begin{abstract}
We study the smallest eigenvalue statistics of the $\beta$-Laguerre and $\beta$-Jacobi ensembles. Using Kaneko's integral formula, we show that the smallest eigenvalue marginal density and distribution functions of the two ensembles for any $\beta>0$ can be represented in terms of multivariate Laguerre and Jacobi polynomials evaluated at a multiple of the identity, provided that the exponent of $x$ in the Laguerre and Jacobi weights is an integer. These representations are readily computable in explicit form using existing symbolic algorithms for multivariate orthogonal polynomials.

From these expressions, we derive new differentiation formulas for the multivariate Laguerre and Jacobi polynomials. Furthermore, we derive explicit solutions to the \plv V and VI differential equations associated with the smallest eigenvalue of the LUE and JUE. We provide numerical experiments and examples. 
\end{abstract}

\maketitle

\section{Introduction}

The $\beta$-Laguerre and $\beta$-Jacobi ensembles are two of the most well-studied random matrix ensembles, also as known as the Wishart and MANOVA matrices. The weight is defined as
\begin{equation}\label{eq:laguerreweight}
    w_{n, \gamma}^{L, \beta}(x_1, \dots, x_n) = \prod_{i<j}|x_i -x_j|^\beta \prod_{i=1}^n x_i^\gamma e^{-x_i/2},
\end{equation}
for the Laguerre, and 
\begin{equation}\label{eq:jacobiweight}
    w_{n, \gamma_1, \gamma_2}^{J, \beta}(x_1, \dots, x_n) = \prod_{i<j}|x_i - x_j|^\beta \prod_{i=1}^nx_i^{\gamma_1}(1-x_i)^{\gamma_2},
\end{equation}
for the Jacobi ensemble, where $n\in\mathbb{N}$, $\gamma, \gamma_1, \gamma_2>-1$ and $\beta>0$. The joint probability densities of the $\beta$-Laguerre ensemble (L$\beta$E) and $\beta$-Jacobi ensemble (J$\beta$E) are given with the weights above with normalizing constants,
\begin{gather}\label{eq:lbedensity}
    f^{L, \beta}_{n, \gamma}(x_1, \dots, x_n) = \mathcal{Z}_{n, \gamma}^{L, \beta}w_{n, \gamma}^{L, \beta}(x_1, \dots, x_n),\hspace{0.5cm} x_i>0, \\ \label{eq:jbedensity}
    f^{J, \beta}_{n, \gamma_1, \gamma_2}(x_1, \dots, x_n) = \mathcal{Z}^{J, \beta}_{n, \gamma_1, \gamma_2}w_{n, \gamma_1, \gamma_2}^{J, \beta}(x_1, \dots, x_n), \hspace{0.5cm}x_i\in[0, 1],
\end{gather}
where the normalization constants are computed by the celebrated Selberg integral \cite{selberg1944bemerkninger},
\begin{gather}\label{eq:laguerreZ}
    \mathcal{Z}^{L, \beta}_{n, \gamma} = 2^{-n(\gamma+1 + \frac{\beta}{2}(n-1))}\prod_{i=0}^{n-1}\frac{\Gamma(1 + \beta/2)}{\Gamma(1 + \beta (i+1)/2)\Gamma(\gamma+1+\beta i/2)}, \\ \label{eq:jacobiZ}
    \mathcal{Z}^{J, \beta}_{n, \gamma_1, \gamma_2} = \prod_{i=0}^{n-1}\frac{\Gamma(1+\beta/2)\Gamma(\gamma_1 + \gamma_2 + \beta(n + i - 1)/2 + 2)}{\Gamma(1+\beta(i+1)/2)\Gamma(1+\gamma_1 + \beta i/2)\Gamma(1+\gamma_2 + \beta i/2)}.
\end{gather}
For $\beta=1,2,4$ these $\beta$-ensembles are often called the orthogonal, unitary, and symplectic ensemble, respectively.

Both the L$\beta$E and J$\beta$E have a hard-edge on their spectra. For the Laguerre ensemble, the eigenvalues cannot be smaller than 0, and for the Jacobi ensemble the eigenvalues cannot be smaller than 0 or/and larger than 1. 
Moreover, as $n\to\infty$ the hard-edge eigenvalue has a limiting distribution, which coincides for the L$\beta$E and the J$\beta$E after appropriate renormalization \cite{nagao1991correlation,forrester1993spectrum,borodin2003increasing}. This limiting distribution is known as the hard-edge scaling limit and is commonly represented by the Fredholm determinant associated with the Bessel kernel \cite{tracy1994level}.

In this work we investigate finite-$n$ hard-edge eigenvalue statistics. In particular, we focus on the density and distribution function of the smallest eigenvalue of the L$\beta$E and J$\beta$E. For finite $n$, the law of the hard-edge eigenvalue admits several representations. We denote by $f_{n,\gamma}^{L,\beta}(x)$ and $F_{n,\gamma}^{L,\beta}(x)$, for $x \in [0,\infty)$, the marginal probability density function and the corresponding cumulative distribution function of the smallest eigenvalue of the L$\beta$E. 

One of the earlier results is the representation of $F_{n, \gamma}^{L, 2}(x)$ as a Fredholm determinant, from which Tracy and Widom derived an expression of $F_{n,\gamma}^{L,2}(x)$ in terms of the solution to the \plv V ordinary differential equation \cite{tracy1994fredholm}. For $\beta=2$ and more generally for $\beta>0$, a number of exact representations are available, including formulas involving hypergeometric functions of a matrix argument and finite dimensional determinants. 

Similarly for the J$\beta$E we also denote by $f_{n, \gamma_1, \gamma_2}^{J, \beta}(x)$ and $F_{n, \gamma_1, \gamma_2}^{J, \beta}(x)$, for $x\in[0, 1]$, the marginal density and distribution function of the smallest eigenvalue. Analogously to the Laguerre case, $F_{n, \gamma_1, \gamma_2}^{J, 2}(x)$ can be represented by a Fredholm determinant, which is further expressed in terms of a solution to a third order ordinary differential equation \cite{tracy1994fredholm}, or a solution to the \plv VI equation \cite{haine1999jacobi}. A detailed list of known exact representations of $f_{n, \gamma}^{L, \beta}(x)$, $F_{n, \gamma}^{L, \beta}(x)$, $f_{n, \gamma_1, \gamma_2}^{J, \beta}(x)$ and $F_{n, \gamma_1, \gamma_2}^{J, \beta}(x)$ can be found in Section~\ref{sec:backgroundlbejbe}; See also \cite{dumitriu2008distributions,forrester2010log,edelman2016beyond,moreno2019extreme,forrester2024computation}.

On the other hand, in 1993, Kaneko \cite{kaneko1993selberg} studied the generalized version of the Selberg integral and Heine's integral formula \cite{heine1878handbuch},
\begin{equation*}
    \int_{[0, 1]^\nu} \prod_{j=1}^\nu \prod_{i=1}^n (x_i - y_j)\prod_{i<j}|x_i - x_j|^\beta \prod_{i=1}^n x_i^{\gamma_1}(1-x_i)^{\gamma_2}dx_1\cdots dx_n. 
\end{equation*}
In the 19th century, Heine studied the $\beta=2$, $\nu=1$ case with a general orthogonal polynomial weight, and noticed that the integral could be expressed as an orthogonal polynomial itself. Selberg \cite{selberg1944bemerkninger} computed the integral without the factor $\prod\prod(x_i- y_j)$ and obtained a product of Gamma functions. Aomoto computed the case $\nu=1$ and found that it can be expressed as a univariate Jacobi polynomial. Kaneko extended this to general $\nu\in\mathbb{N}$, expressing the integral in terms of the hypergeometric function of a matrix argument, and further proved that it can also be written as a multivariate Jacobi polynomial. This result can be thought of as a full multivariate generalization of Heine's integral formula.

Suppose that $\gamma, \gamma_1$ are in $\mathbb{N}$. Our main result uses Kaneko's integral formula to show that the functions $F_{n, \gamma}^{L, \beta}(x)$, $F_{n, \gamma}^{L, \beta}(x)$, $F_{n, \gamma_1, \gamma_2}^{J, \beta}(x)$, and $f_{n, \gamma_1, \gamma_2}^{J, \beta}(x)$ can be expressed exactly in terms of a scaled multivariate Laguerre or Jacobi polynomial evaluated at the diagonal, i.e., $P(x_1, \dots, x_m)|_{x_1=\cdots =x_m}$, where $P$ is a multivariate Laguerre or Jacobi polynomial. We believe this is among the most efficient methods for obtaining explicit rational expressions of these functions, as the symbolic forms of multivariate Laguerre and Jacobi polynomials can be computed easily using existing algorithms (e.g., Gram-Schmidt orthogonalization) and software \cite{dumitriu2007mops}. We note that in the J$\beta$E case, the same result for the distribution function $F_{n, \gamma_1, \gamma_2}^{J, \beta}(x)$ is reported in a recent work by Winn \cite{winn2024extreme}. We further note that the level densities of the L$\beta$E and J$\beta$E have been derived for even $\beta\in\mathbb{N}$ by Baker and Forrester using an analogous method \cite{baker1997calogero}.

From our newly obtained representations and the elementary relationship between the density and distribution functions, we derive new differentiation formulas for the multivariate Laguerre and Jacobi polynomials. Differentiation formulas for multivariate Laguerre polynomials were first obtained in \cite[Section 4.2]{baker1997calogero} for general partition indices. Our formula specializes to the square partition indices, and applies to the univariate function arising from multivariate polynomials evaluated at a multiple of the identity. As a consequence, we obtain explicit rational solutions of the \plv V and VI equations associated with the smallest eigenvalue of the LUE and JUE. 

We summarize our contribution as follows.
\begin{itemize}
    \item We obtain new exact representations of the smallest eigenvalue marginal density function (PDF) and distribution function (CDF) of the L$\beta$E and J$\beta$E in terms of multivariate Laguerre and Jacobi polynomials. (Theorems~\ref{thm:laguerremainresult},~\ref{thm:jacobimainresult})
    \item We derive new differentiation formulas for the multivariate Laguerre and Jacobi polynomials (Corollaries~\ref{cor:laguerrediff},~\ref{cor:jacobidiff}). Moreover, we found explicit solutions of the \plv V, VI equations associated to the LUE and JUE smallest eigenvalues (Corollaries~\ref{cor:laguerreplv},~\ref{cor:jacobiplv}). 
\end{itemize}

The remainder of the paper is organized as follows. Section~\ref{sec:background} reviews background material on the L$\beta$E, J$\beta$E, and known exact representations of the hard-edge eigenvalue, as well as a brief overview of multivariate statistics and multivariate orthogonal polynomials. In Section~\ref{sec:Laguerremainresult}, we present new representations of the smallest eigenvalue statistics of the L$\beta$E, with analogous results for the J$\beta$E given in Section~\ref{sec:jacobimainresult}. Section~\ref{sec:numerics} provides a short numerical simulation to verify our results.

\section{Background}\label{sec:background}

\subsection{Multivariate orthogonal polynomials}

In this section, we briefly review the basic definitions and properties of functions in multivariate statistics for general $\beta>0$. We also recall the definitions of the multivariate orthogonal polynomials, namely the Jack polynomial, the multivariate Laguerre polynomial, and the multivariate Jacobi polynomial. For more details please refer to, for instance, \cite{baker1997calogero,dumitriu2007mops,forrester2010log,macdonald1998symmetric}.

\subsubsection{Basic definitions}

Throughout this work, we will use the $\beta$ parameter, as opposed to the parameter $\alpha = 2/ \beta$. We first start by defining the partition:
\begin{definition}
    A partition $\kappa=[\kappa_1, \dots, \kappa_n]$ is a non-increasing ordered sequence of positive integers. If $k = \sum_{i=1}^n \kappa_i$, we say that $\kappa$ is a partition of $k$ and denote by $\kappa\vdash k$. Also we write $l(\kappa) = n$. 
\end{definition}
Recall the Pochhammer symbol defined with the Gamma function
\begin{equation*}
    (a)_k = \frac{\Gamma(a+k)}{\Gamma(a)}. 
\end{equation*}
A multivariate extension of the Gamma function and Pochhammer symbol are defined as follows. For a partition $\kappa = [\kappa_1, \dots, \kappa_n]$ such that $\kappa\vdash k$,
\begin{equation}
    (a)_\kappa^\beta := \prod_{i=1}^n \left(a - \frac{\beta(i-1)}{2}\right)_{\kappa_i},\hspace{1cm}\Gamma_n^{\beta}(a) := \pi^{\frac{\beta n(n-1)}{4}}\prod_{i=1}^n \Gamma\left(a - \frac{\beta (i-1)}{2}\right).
\end{equation}

\subsubsection{Jack polynomials}

Jack polynomials are often referred to as ``multivariate monomials." In the case $\beta=1$, they appear as the classical zonal polynomials, extensively studied in multivariate statistics \cite{constantine1963some,james1961zonal,muirhead1982aspects}, where they arise as eigenfunctions of the Laplace–Beltrami operator on symmetric spaces, in particular on the cone of real positive definite matrices. For $\beta=2$, Jack polynomials are called Schur polynomials, and they frequently appear in combinatorics and representation theory \cite{macdonald1998symmetric,stanley1999enumerative}. 

A Jack polynomial in $n$ variables is indexed by a partition $\kappa$ with $l(\kappa)\leq n$. The leading term of the Jack polynomial indexed by $\kappa$ is the symmetric sum of $x_{i_1}^{\kappa_1}\cdots x_{i_{l(\kappa)}}^{\kappa_{l(\kappa)}}$. (Here $i_1, \dots, i_{l(\kappa)}$ are distinct and we use dominance order of partitions to determine the highest-order term.) The Jack polynomial in $n$ variables indexed by $\kappa$ is an eigenfunction of the differential operator 
\begin{equation*}
    \sum_{i=1}^n x_i^2 \frac{d^2}{dx_i^2} + \beta \sum_{i\neq j} \frac{x_i^2}{x_i - x_j}\frac{d}{dx_i},
\end{equation*}
associated to the eigenvalue $\sum_{i=1}^n\left(\kappa_i(\kappa_i-1-\beta(i-1)/2)+ \beta k(n-1)\right)$. 

There are three frequently used normalizations for the Jack polynomials, often denoted by P, J, and C-normalization. In this work, we use the C-normalization with two parameters $\beta>0$ and $\kappa\vdash k$, written as $C_\kappa^{\beta}$, which satisfies the following identity
\begin{equation*}
    (x_1 + \dots+ x_n)^k = \sum_{\kappa \vdash k, l(\kappa)\leq n} C_\kappa^\beta(x_1, \dots, x_n).  
\end{equation*}
As with all other multivariate orthogonal polynomials, they take a matrix argument, i.e., $C_\kappa^\beta(X) := C_\kappa^\beta(x_1, \dots, x_n)$ if the eigenvalues of $X$ are $x_1, \dots, x_n$. For more on Jack polynomials, see for instance \cite{baker1997calogero,kaneko1993selberg,stanley1989some,stanley1999enumerative}.

\subsubsection{Hypergeometric function of a matrix argument}
The Hypergeometric functions of a matrix argument are also called the multivariate hypergeometric functions. It is defined as
\begin{equation}
    {}_pF_q^{\beta}(a_1, \dots, a_p; b_1, \dots, b_q; x_1, \dots, x_n) = \sum_{k=0}^\infty \sum_{\kappa\vdash k}\frac{(a_1)_\kappa^\beta\cdots(a_p)_\kappa^\beta}{k!(b_1)_\kappa^\beta\cdots (b_q)_\kappa^\beta} C_\kappa^\beta(x_1, \dots, x_n),
\end{equation}
where $C_\kappa^\beta$ is the Jack polynomial with C-normalization. We define
\begin{equation*}
    {}_pF_q^{\beta}(a_1, \dots, a_p; b_1, \dots, b_q; X):={}_pF_q^{\beta}(a_1, \dots, a_p; b_1, \dots, b_q; x_1, \dots, x_n),
\end{equation*}
where $X$ is an $n\times n$ matrix with eigenvalues $x_1, \dots, x_n$. Multivariate hypergeometric function is used in numerous representations in multivariate statistics, including the eigenvalues of random matrices. For computing hypergeometric functions of a matrix argument see \cite{koev2006efficient}.

\subsubsection{Multivariate Laguerre polynomials}

A multivariate Laguerre polynomial is defined as a multivariate symmetric polynomial indexed by a partition associated with the weight $w_{n, \gamma}^{L, \beta}(x_1, \dots, x_n)$ defined in \eqref{eq:laguerreweight}. With parameters $\beta>0, \gamma>-1, n\in\mathbb{N}$, we denote by
\begin{equation*}
    L_{\kappa, \gamma}^\beta(x_1, \dots, x_n),
\end{equation*}
the multivariate Laguerre polynomial indexed by the partition $\kappa$. These polynomials are orthogonal with respect to the weighted inner product
\begin{equation*}
    \langle L_{\kappa, \gamma}^{\beta}, L_{\mu, \gamma}^{\beta} \rangle := \int_{(\mathbb{R}_+)^n} L_{\kappa, \gamma}^{\beta}(x_1, \dots, x_n)L_{\mu, \gamma}^{\beta}(x_1, \dots, x_n)dx_1\cdots dx_n.
\end{equation*}
They are also written as functions of a square matrix, i.e., $L_{\kappa, \gamma}^{\beta}(X):=L_{\kappa, \gamma}^{\beta}(x_1, \dots, x_n)$, where $x_1, \dots, x_n$ are the eigenvalues of the $n\times n$ matrix $X$. In the next section we will often evaluate multivariate Laguerre polynomials at a multiple of the identity, $X = g(x)I_n$. 

Multivariate Laguerre polynomials first appeared for $\beta=1, 2$ in the literature, for instance, \cite{james1964distributions,muirhead1982aspects}. For general $\beta>0$ they appear in \cite{lassalle1991laguerre} and were further studied in \cite{baker1997calogero}. 

Contrary to the Jack polynomials which have popular choices of normalization, multivariate Laguerre (and also Jacobi) polynomials have a number of different normalizations that are used in the literature. We take the following normalization with respect to the above inner product
\begin{align*}
    \langle L_{\kappa, \gamma}^{\beta}, L_{\mu, \gamma}^{\beta} \rangle = \delta_{\kappa \mu} 2^{2|\kappa|+n(1+\gamma+\beta(n-1)/2)} \frac{(|\kappa|!)^2}{\beta^{2|\kappa|} j(\kappa; \beta)} \left(\frac{\beta n}{2}\right)_\kappa^{\beta}&\left(1 + \gamma + \frac{\beta(n-1)}{2}\right)^{\beta}_\kappa \\
    &\times \frac{\Gamma_n^\beta\left(1+\frac{\beta n}{2}\right)\Gamma_n^\beta\left(1+\gamma+\frac{\beta (n-1)}{2}\right)}{\Gamma\left(1+\frac{\beta}{2}\right)^n}.
\end{align*}
The function $j(\kappa;\beta)$ is defined as a product of upper and lower hook lengths of the partition $\kappa$ 
\begin{equation*}
    j(\kappa;\beta) = \prod_{s\in\kappa} \left(l_{\kappa}(s)+\frac{2}{\beta}(1+a_\kappa(s))\right)\left(l_\kappa(s) + 1+\frac{2}{\beta}a_\kappa(s)\right),
\end{equation*}
where $a_\kappa(s)$ and $l_\kappa(s)$ denote the arm- and leg-lengths of the box $s \in \kappa$, respectively, in the Ferrers diagram of $\kappa$. See \cite[Equation (2.10a)]{baker1997calogero} or \cite[Definition 2.14]{dumitriu2007mops} for details. This normalization corresponds to the change of variables $x_i \mapsto x_i/2$ in the multivariate Laguerre polynomials of \cite{dumitriu2002matrix}. We provide some examples of the multivariate Laguerre polynomials in Appendix~\ref{sec:examples} for reference. 

Finally, the constant term of a multivariate Laguerre polynomial $L_{n, \gamma}^\beta$ is given by
\begin{equation}\label{eq:zerolaguerre}
    L_{\kappa, \gamma}^{\beta}(0, \dots, 0) = \left(\frac{2}{\beta}\right)^{2|\kappa|}\frac{|\kappa|!}{j(\kappa;\beta)}\left(\frac{\beta n}{2}\right)_\kappa^{\beta}\left(1 + \gamma + \frac{\beta(n-1)}{2}\right)^{\beta}_\kappa.
\end{equation}

\subsubsection{Multivariate Jacobi polynomial}

A multivariate Jacobi polynomial is another multivariate symmetric polynomial indexed by a partition, which is orthogonal with respect to the weighted inner product
\begin{equation*}
    \langle f, g\rangle = \int_{[0, 1]^n}f(x_1, \dots, x_n)g(x_1, \dots, x_n)w_{n, \gamma_1, \gamma_2}^{J, \beta}(x_1, \dots, x_n)dx_1 \cdots dx_n.
\end{equation*}
The multivariate Jacobi polynomial with $n$ variables indexed by the partition $\kappa$ is denoted by
\begin{equation*}
    P_{\kappa, \gamma_1, \gamma_2}^\beta(x_1, \dots, x_n),
\end{equation*}
and can also be expressed with a matrix argument $X$ with eigenvalues $x_1, \dots, x_n$ similarly to the Laguerre case. For the Jacobi polynomials, we take the following simple normalization, 
\begin{equation*}
    P_{\kappa, \gamma_1, \gamma_2}^\beta (0_n) = 1. 
\end{equation*}
The multivariate Jacobi polynomials are eigenfunctions of the radial Laplace-Beltrami operator (spherical function) on Grassmannians, for $\beta=1,2,4$. More precisely, they appear as spherical functions of the symmetric spaces $\text{O}(n)/\text{O}(p)\times \text{O}(n-p)$, $\text{U}(n)/\text{U}(p)\times \text{U}(n-p)$, and $\text{USp}(n)/\text{USp}(p)\times \text{USp}(n-p)$. James and Constantine \cite{james1974generalized} considered the real and complex ($\beta=1,2$) cases, and further generalized to general $\beta>0$ in \cite{koornwinder1974orthogonal} for two variables and for arbitrary number of variables in \cite{lassalle1991jacobi}. For general $\beta>0$, the root system can be analyzed through Dunkl operators \cite{dunkl1988reflection}. Further properties for general $\beta>0$ were investigated in \cite{baker1997calogero}. We also provide some examples of the multivariate Jacobi polynomials $P_{\kappa, \gamma_1, \gamma_2}^\beta(x_1, \dots, x_n)$ in Appendix~\ref{sec:examples}.

\subsection{Exact representations for the smallest eigenvalue of the L$\beta$E and J$\beta$E}\label{sec:backgroundlbejbe}

\subsubsection{Matrix models for Laguerre and Jacobi ensembles}

Consider an $m\times n$ matrix $A$ filled with i.i.d. real Gaussian variables $(m\geq n)$. The eigenvalues of the matrix $A^TA$, which is often called the \textit{real Wishart matrix}, are equivalent to the Laguerre Orthogonal Ensemble (LOE) and were considered in the statistics literature in the early 20th century. Replacing the real Gaussians with complex or quaternionic Gaussian variables, one obtains the Laguerre Unitary Ensemble (LUE, $\beta=2$) and the Laguerre Symplectic Ensemble (LSE, $\beta=4$). More precisely, these ensembles correspond to the L$\beta$E with the parameter 
\begin{equation*}
    \gamma = \frac{\beta}{2}(m - n + 1) - 1.
\end{equation*} 
Similarly, the Jacobi Orthogonal Ensemble (JOE) is obtained by taking the eigenvalues of the matrix $A^TA/ (A^TA + B^TB)$ constructed from two Wishart matrices $A^TA, B^TB$, which is sometimes called the \textit{double Wishart matrix}. Moreover, the JUE and the JSE are obtained by replacing the real Gaussian variables with complex and quaternionic Gaussian variables (and also replacing the real transpose by the conjugate transpose). The matrix $A$ is $m_1\times n$ and the matrix $B$ is $m_2\times n$ with $m_1, m_2\geq n$. The parameters $\gamma_1, \gamma_2>-1$ are then given by 
\begin{equation*}
    \gamma_1 = \frac{\beta}{2}(m_1 - n + 1) - 1, \hspace{0.5cm} \gamma_2 = \frac{\beta}{2}(m_2 - n + 1) - 1,
\end{equation*}
for given $m_1, m_2, n, \beta$. 

The joint eigenvalue densities of the LOE, LUE, and LSE are explicitly computable and correspond to the cases $\beta=1,2,4$ of \eqref{eq:lbedensity}; analogous formulas hold for the Jacobi ensembles. These densities extend naturally to general $\beta>0$, although no explicit $\beta$-dimensional Gaussian matrix exists in this case. Dumitriu and Edelman \cite{dumitriu2002matrix} introduced a sparse tridiagonal matrix model whose eigenvalue distribution coincides with that of the L$\beta$E. This construction may be viewed heuristically as the outcome of applying a numerical linear algebra routine to a hypothetical $\beta$-dimensional Wishart matrix. Subsequently, several matrix models for the J$\beta$E were proposed in \cite{lippert2003matrix,killip2004matrix,edelman2008beta}.

\subsubsection{The smallest eigenvalue of the Laguerre ensemble}

The smallest eigenvalue of the LOE has long been of central importance in statistics; see \cite{muirhead1982aspects} and the references therein for background. An early result for its density is the expression in terms of a hypergeometric function of matrix argument due to \cite{krishnaiah1971exact}:
\begin{equation*}
    f_{n, \gamma}^{L, 1}(x)  = x^{n\gamma} e^{-\frac{xn}{2}} {}_2F_0(-\gamma, 1+n/2; -2 I_{n-1}/x).
\end{equation*}
This was later extended to general $\beta>0$ in \cite[Eq.~(2.12b)]{forrester1994exact} and \cite{dumitriuthesis}. The distribution function was likewise expressed in terms of a hypergeometric function of a matrix argument for arbitrary $\beta>0$ in \cite[Eq.~(2.13a)]{forrester1994exact}, using Kaneko's integral formula. A related representation in terms of the sum of Jack polynomials was obtained in \cite{edelman2014eigenvalue}, adapting techniques discussed in \cite{muirhead1982aspects}.

For $\beta=2$, the cumulative distribution of the smallest eigenvalue of the LUE admits the Fredholm determinant representation
\begin{equation*}
    1 - F_{n, \gamma}^{L, 2}(x) = \mathbb{P}(\text{No eigenvalue in }(0, x)) = \det(I - K_{n, \gamma}^L|_{L^2([0, x])}),
\end{equation*}
where $K^L_{n, \gamma}$ denotes the Laguerre kernel. This formula can be extended to $\beta=1,4$ using the ($2\times 2$ matrix-valued) determinantal expressions for $\prod_{i<j}|x_i - x_j|^\beta$, which reduce to Fredholm Pfaffian formulas in both cases \cite{mehta2004random}. 

In 1994 Tracy and Widom~\cite{tracy1994fredholm} introduced an important representation of $F_{n, \gamma}^{L, 2}(x)$
\begin{equation}\label{eq:LUEpainleve}
    1 - F_{n, \gamma}^{L, 2}(x) = \exp\left(-\int_0^{x/2} \frac{\sigma(s)}{s}ds\right),
\end{equation}
where $\sigma(x) = \sigma_{V}(x)$ is the solution of the fifth \plv ordinary differential equation,
\begin{equation}\label{eq:painleve5}
    (x\sigma'')^2 = 4x(\sigma')^3 + \sigma^2 + (2\gamma + 4n - 2x) \sigma \sigma' + (\gamma^2 - 2\gamma x - 4nx + x^2) (\sigma')^2 - 4\sigma (\sigma')^2,
\end{equation}
subject to the asymptotic condition
\begin{equation}\label{eq:painleve5cond}
    \sigma(x)\simeq \frac{\Gamma(n+\gamma+1)}{\Gamma(n)\Gamma(\gamma+1)\Gamma(\gamma+2)}x^{\gamma+1}. 
\end{equation}

There are also finite dimensional determinantal formulas for $F_{n, \gamma}^{L, 2}$. For $\gamma\in\mathbb{N}$, Forrester and Hughes obtained a $\gamma\times \gamma$ determinant involving univariate Laguerre polynomials \cite[Eq.~(3.19)]{forrester1994complex}. An $n\times n$ determinant representation with a matrix of gamma functions also appears in \cite{edelman2016beyond}. In another direction, a recurrence relation that computes $f_{n, \gamma+1}^{L, \beta}$ from $f_{n, 0}^{L, \beta}, \dots, f_{n, \gamma}^{L, \beta}$ is introduced in Edelman \cite{edelman1989thesis,edelman1991distribution} for $\beta=1$. This is generalized to $\beta>0$ in \cite{kumar2019recursion}. Within this recurrence framework, it follows that the density of the smallest eigenvalue of the L$\beta$E has the form 
\begin{equation*}
    f_{n, \gamma}^{L, \beta}(x) = C x^\gamma e^{-\frac{\beta n x}{2}} P(x),
\end{equation*}
where $C$ is a constant and $P$ is a polynomial of degree  $\gamma(n-1)$. 

\subsubsection{The smallest eigenvalue of the Jacobi ensemble}

In this work, we restrict our attention to the smallest eigenvalue of the Jacobi ensembles, noting that the largest eigenvalue can be obtained from it via the parameter exchange $(\gamma_1,\gamma_2)\mapsto(\gamma_2,\gamma_1)$.

Because of the close relationship between the Laguerre and Jacobi ensembles, many of the results for the Laguerre case have corresponding analogues for the Jacobi ensembles. The largest eigenvalue cumulative distribution of the JOE is given in terms of the hypergeometric function of a matrix argument \cite[Eq.~(61)]{constantine1963some}. Using again Kaneko’s integral formula \cite{kaneko1993selberg}, Dumitriu and Koev \cite[Theorem~3.1]{dumitriu2008distributions} extended the $\beta=1$ case to arbitrary $\beta>0$
\begin{align*}
    F_{n, \gamma_1, \gamma_2}^{J, \beta}(x) &= 1 - \prod_{i=0}^{n-1}\frac{\Gamma(\gamma_1+\gamma_2+2+\beta(n-1))\Gamma(\beta(n-1-i)/2+1)}{\Gamma(\gamma_2+2+\beta(2n-2-i)/2)\Gamma(\gamma_1+1+\beta(n-i-1)/2)}\\ 
    \times (1-&x)^{n(\gamma_1 + 1 + \beta(n-1)/2)} \times {}_2 F_1^{(\beta)} \left(\gamma_2 + 1 + \frac{\beta}{2}(n-1), -\gamma_1; \gamma_2 + 2+ \beta(n-1); (1-x)I_n\right).
\end{align*}
The density function for general $\beta>0$ is also given in terms of the hypergeometric function of a matrix argument in \cite{dumitriu2012smallest}. When $\gamma_1 \in \mathbb{N}$, this hypergeometric function reduces to a finite sum of Jack polynomials \cite[Eq.~(3.3)]{dumitriu2008distributions}. This observation was further investigated in \cite{forrester2024computation}, where a recursion scheme for computing $F_{n,\gamma_1,\gamma_2}^{J,\beta}$ was developed.

For the JUE, the distribution of the smallest eigenvalue has the Fredholm determinant expression
\begin{equation*}
    1 - F_{n, \gamma_1, \gamma_2}^{J, 2}(x) = \det(I - K_{n, \gamma_1, \gamma_2}^{J}|_{L^2([0, x])}),
\end{equation*}
where $K_{n,\gamma_1,\gamma_2}^J$ denotes the finite Jacobi kernel \cite{tracy1994fredholm}. Fredholm Pfaffian formulas exist for $\beta=1, 4$ cases, analogously to the Laguerre ensembles \cite{mehta2004random}. Tracy and Widom \cite[Eq.~5.62]{tracy1994fredholm} derived a third-order differential equation from the Fredholm determinant representation, using the method analogous to that leading to the \plv V equation in the Laguerre case. However, this equation is not known to reduce to second order. An alternative derivation was obtained by Haine and Semengue \cite{haine1999jacobi} using Virasoro constraints
\begin{equation*}
    1-F_{n,\gamma_1, \gamma_2}^{J, 2}(x) = \exp\left(-\int_0^x \frac{\sigma(s)}{s(1-s)}ds\right),
\end{equation*}
where $\sigma$ is a solution to a \plv VI equation shifted by a linear function. Other indirect methods that connect $F_{n, \gamma_1, \gamma_2}^{J, 2}$ to a solution of a \plv VI equation have been investigated, for instance, in \cite{duenez2010lowest}.
Finite-dimensional determinant formulas exist for $F_{n, \gamma_1, \gamma_2}^{J, 2}$, an $n\times n$ determinant \cite{khatri1964distribution} and a $(\gamma_1+\gamma_2)\times (\gamma_1+\gamma_2)$ determinant was proposed in \cite{moreno2019extreme} for $\gamma_1, \gamma_2\in\mathbb{N}$. Recently, in \cite{winn2024extreme}, the cumulative distribution function $F_{n, \gamma_1, \gamma_2}^{J, \beta}$ was expressed in terms of a multivariate Jacobi polynomial.

\section{Laguerre ensemble case}

For $\beta>0$ and $\gamma>-1$, we introduce the dual parameters \cite{forrester2025dualities}
\begin{equation}\label{eq:dualparam}
    \tilde{\beta} :=\frac{4}{\beta}, \hspace{0.5cm}\tilde{\gamma} := \frac{2}{\beta}(\gamma+1)-1,
\end{equation}
that appear frequently in the following sections. Moreover, let us denote by $n^{(m)}$ the partition of $nm$ with $m$ copies of $n$, 
\begin{equation}\label{eq:squarepartitiondef}
    n^{(m)} := [\,\,\underbrace{n, \,\dots, \,n}_{m}\,\,] \vdash nm.
\end{equation}

\subsection{Marginal density and distribution of the smallest eigenvalue of the Laguerre ensemble}\label{sec:Laguerremainresult}

Taking the limit of Jacobi polynomials to Laguerre polynomials, we obtain the following Laguerre-analogue of Kaneko's integral formula \cite{kaneko1993selberg}
\begin{lemma}\label{lem:kanedolaguerre}
    For $n, \nu\in\mathbb{N}, \beta>0, \gamma>-1$, we have
    \begin{equation}\label{eq:kanekolaguerre}
        \int_{\mathbb{R}_+^n} \prod_{j=1}^\nu\prod_{i=1}^n (x_i - y_j) w_{n, \gamma}^{L, \beta}(x_1, \dots, x_n) dx_1\cdots dx_n = D_{n, \gamma}^{L, \beta} \times L_{n^{(\nu)}, \tilde{\gamma}}^{\tilde{\beta}}\bigg(\frac{2y_1}{\beta}, \dots, \frac{2y_\nu}{\beta}\bigg),
    \end{equation}
    where the constant term is
    \begin{equation*}
        D_{n, \gamma}^{L, \beta} = \frac{1}{\mathcal{Z}_{n, \gamma+\nu}^{L, \beta} \cdot L_{n^{(\nu)}, \tilde{\gamma}}^{\tilde{\beta}}(0_\nu)},
    \end{equation*}
    and $\tilde{\beta}, \tilde{\gamma}$ is defined in \eqref{eq:dualparam}. 
\end{lemma}
This integral appeared in \cite{baker1997calogero} and also in \cite{luque2003hankel} along with other limits of the Jacobi polynomials. The hypergeometric function analogue of \eqref{eq:kanekolaguerre} was also discussed in both references. Recall that the value of the multivariate Laguerre polynomial evaluated at $0$ can be found above in \eqref{eq:zerolaguerre}. 

As an immediate consequence of Lemma~\ref{lem:kanedolaguerre} we derive an identity that can be used to reduce the number of variables for multivariate Laguerre polynomials indexed by square partitions. By moving some of the $x_i$'s to $(x_i - 0)$'s, we get the following result. 
\begin{proposition}\label{prop:integrationconsq}
    For any $a, n, \nu\in\mathbb{N}$, $\beta>0$, $\gamma>-1$, we have
    \begin{equation*}
        L_{n^{(\nu)}, \gamma}^\beta(x_1, \dots, x_\nu) \propto L_{n^{(\nu+a)}, \gamma-\frac{\beta a}{2}}^\beta (x_1, \dots, x_\nu, \underbrace{0, 0, \dots, 0}_{a}),
    \end{equation*}
    where the ratio can be obtained by comparing the constant terms in \eqref{eq:zerolaguerre}. 
\end{proposition}

Let us now consider the cumulative distribution of the smallest eigenvalue of the L$\beta$E. From \eqref{eq:lbedensity}, the distribution of the smallest eigenvalue of L$\beta$E is 
\begin{equation*}
    F_{n, \gamma}^{L, \beta}(x) = \mathbb{P}(x < x_{\min{}}) = 1 - \mathbb{P}(\text{all eigenvalues in }[x,\infty)),
\end{equation*}
and we further obtain
\begin{equation*}
    1-F_{n, \gamma}^{L, \beta}(x) = \mathcal{Z}_{n, \gamma}^{L,\beta}\int_{[x, \infty)^n} \prod_{1\leq i < j \leq n} |x_i - x_j|^\beta \prod_{i=1}^n e^{-x_i/2}x_i^\gamma dx_1\cdots dx_n.
\end{equation*}
Suppose that $\gamma\in\mathbb{Z}$. Using the change of variables $t_i = x_i - x$ and Lemma~\ref{lem:kanedolaguerre} we get
\begin{align}\nonumber
    1-F_{n, \gamma}^{L, \beta}(x) &= \mathcal{Z}_{n, \gamma}^{L, \beta}e^{-nx/2} \int_{\mathbb{R}_+^n}\prod_{1\leq i < j \leq n} |t_i - t_j|^\beta \prod_{i=1}^n e^{-t_i/2}(t_i+x)^\gamma dt_1\cdots dt_n\\ \label{eq:lbedistpoly}
    &=  \frac{1}{{L_{n^{(\gamma)}, \frac{2}{\beta}-1}^{\tilde{\beta}}(0_\gamma)}}\times e^{-nx/2} L_{n^{(\gamma)}, \frac{2}{\beta}-1}^{\tilde{\beta}}\left(-\frac{2x}{\beta} I_\gamma\right),
\end{align}
where the constant term is given as
\begin{equation}\label{eq:lbedistconstantsimple}
    L_{n^{(\gamma)}, \frac{2}{\beta}-1}^{\tilde{\beta}}(0_\gamma) = (\gamma n)!\prod_{i=0}^{\gamma-1}\frac{\Gamma(n+2(i+1)/\beta)\Gamma(1+2i/\beta)}{\Gamma(2(i+1)/\beta)\Gamma(n+1+2i/\beta)},
\end{equation}
and $n^{(\gamma)} = [n, \cdots, n]\vdash n\gamma$ is a partition of $\gamma n$ with $\gamma$ copies of $n$, as defined in \eqref{eq:squarepartitiondef}. 

To compute the marginal density of the smallest eigenvalue of L$\beta$E, we compute the integral
\begin{align*}
    f_{n, \gamma}^{L, \beta}(x) &= n\int_{[x, \infty)^{n-1}}f_{n, \gamma}^{L, \beta}(x_1, \dots, x_{n-1}, x) dx_1\cdots dx_n\\
    &= n \mathcal{Z}_{n, \gamma}^{L, \beta} x^\gamma e^{-\frac{nx}{2}}\int_{\mathbb{R}_+^{n-1}}w_{n-1, \beta}^{L, \beta}(t_1, \dots, t_{n-1}) \prod_{i=1}^{n-1} (t_i + x)^\gamma dt_1\cdots dt_{n-1},
\end{align*}
once more by the change of variables $t_i= x_i - x$ for all $i$. From Lemma~\ref{lem:kanedolaguerre} we get
\begin{equation*}
    f_{n, \gamma}^{L, \beta}(x) = \frac{n\mathcal{Z}_{n, \gamma}^{L, \beta}}{\mathcal{Z}_{n-1, \beta+\gamma}^{L, \beta}\cdot L_{(n-1)^{(\gamma)}, \frac{2}{\beta}+1}^{\tilde{\beta}}(0)} \times x^\gamma e^{-\frac{nx}{2}} L_{(n-1)^{(\gamma)}, \frac{2}{\beta}+1}^{L, \tilde{\beta}}\left(-\frac{2x}{\beta} I_\gamma\right).
\end{equation*}
Summarizing the two representations we get the following result. 

\begin{theorem}\label{thm:laguerremainresult}
    Let $\beta>0$ and suppose that $\gamma>-1$ is an integer. The smallest eigenvalue density and distribution of the $n\times n$ L$\beta$E with parameter $\gamma$ can be written in terms of the multivariate Laguerre polynomial indexed by square partitions,
    \begin{gather*}
        F_{n, \gamma}^{L, \beta}(x) = 1 - \frac{1}{{L_{n^{(\gamma)}, \frac{2}{\beta}-1}^{\tilde{\beta}}(0_\gamma)}}\times e^{-\frac{nx}{2}} L_{n^{(\gamma)}, \frac{2}{\beta}-1}^{\tilde{\beta}}\left(-\frac{2x}{\beta} I_\gamma\right),\\
        f_{n, \gamma}^{L, \beta}(x) = \frac{n\mathcal{Z}_{n, \gamma}^{L, \beta}}{\mathcal{Z}_{n-1, \beta+\gamma}^{L, \beta}\cdot L_{(n-1)^{(\gamma)}, \frac{2}{\beta}+1}^{\tilde{\beta}}(0)} \times x^\gamma e^{-\frac{nx}{2}} L_{(n-1)^{(\gamma)}, \frac{2}{\beta}+1}^{\tilde{\beta}}\left(-\frac{2x}{\beta} I_\gamma\right),
    \end{gather*}
    where the constants can be computed by \eqref{eq:laguerreZ} and \eqref{eq:zerolaguerre}.
\end{theorem}

The validity of Theorem~\ref{thm:laguerremainresult} may be confirmed by comparison with previously reported results. In Table 1 of \cite{kumar2019recursion}, explicit expressions of the marginal density of the L$\beta$E has been computed. For instance, with $\beta=\frac{1}{2}$, $n=4$, $\gamma=3$, we have
\begin{equation*}
    \frac{Z_{n, \gamma}^{L, \beta}}{Z_{n-1, \beta+\gamma}^{L, \beta}} = \frac{16}{585}, \hspace{1cm} L_{(n-1)^{(\gamma)}, \frac{2}{\beta}+1}^{\tilde{\beta}}(0) = 26238643200,
\end{equation*}
and $L_{(n-1)^{(\gamma)}, \frac{2}{\beta}+1}^{\tilde{\beta}}(-\frac{2x}{\beta} I_\gamma)$ is listed in \eqref{eq:kumarcomparison}. From Theorem~\ref{thm:laguerremainresult} we can write down the explicit expression of the probability density of the smallest eigenvalue 
\begin{align*}
    f_{4, 3}^{L, \frac{1}{2}}(x) = \frac{8}{212837625} e^{-2x} x^3 (x^9 + 36x^8 +& 630x^7 + 6846x^6 + 50274x^5 + 259308x^4 \\
    &+ 941094x^3 + 2345112x^2 + 3742200x + 2910600),
\end{align*}
which agrees with the first entry of \cite[Table 1]{kumar2019recursion} after the change of variable $x\mapsto \beta x=x/2$ due to a different normalization of the L$\beta$E. Integrating the above density we get
\begin{align*}
    \int_0^xf_{4, 3}^{L, \frac{1}{2}}(s)ds &= 1-\frac{e^{-2x}}{212837625} ( 4x^{12} + 168x^{11} + 3444x^{10} + 44604x^{9} +  01814x^{8}\\ 
    &+ 2644488x^{7} + 13020084x^{6} + 48440700x^{5} + 136070550x^{4} + 283783500x^{3} \\
    &+ 425675250x^{2} + 425675250x + 212837625 ),
\end{align*}
which agrees with $F_{4, 3}^{L, \frac{1}{2}}$ obtained by Theorem~\ref{thm:laguerremainresult} with $L_{[4,4,4],3}^8(-4x, -4x, -4x)$ (written in \eqref{eq:laguerresectionexF}). 

\subsection{Differentiation formula and explicit \plv V solution}

To derive a differentiation formula, we consider the elementary relationship between the distribution function and the density function. We obtain the following differentiation formula for multivariate Laguerre polynomials at the multiple of the identity,
\begin{align*}
    \frac{d}{dx}&L_{n^{(\gamma)}, \frac{2}{\beta}-1}^{\tilde{\beta}}(x I_\gamma) \\
    &= \frac{(-1)^\gamma n\mathcal{Z}_{n, \gamma}^{L, \beta}\cdot L_{n^{(\gamma)}, \frac{2}{\beta}-1}^{\tilde{\beta}}(0_\gamma)}{\mathcal{Z}_{n-1, \beta+\gamma}^{L, \beta}\cdot L_{(n-1)^{(\gamma)}, \frac{2}{\beta}+1}^{\tilde{\beta}}(0_\gamma)}\left(\frac{\beta}{2}\right)^{\gamma+1}x^\gamma L_{(n-1)^{(\gamma)}, \frac{2}{\beta}+1}^{\tilde{\beta}}(x I_\gamma)-\frac{\beta n}{4} L_{n^{(\gamma)}, \frac{2}{\beta}-1}^{\tilde{\beta}}(x I_\gamma).
\end{align*}
Changing the parameter $\beta\to\tilde{\beta}$ and simplifying the constants, we obtain the following result. We remark that basic differentiation formulas for multivariate Laguerre polynomials can be found in \cite[Eq.~(4.17a), (4.17b)]{baker1997calogero}. 

\begin{corollary}\label{cor:laguerrediff}
    For $\beta>0$, $n \in\mathbb{N}$ and a nonnegative integer $\gamma>-1$, the multivariate Laguerre polynomial satisfies the following differentiation formula,
    \begin{equation}\label{eq:laguerrediff}
        \frac{d}{dx}L_{n^{(\gamma)}, \frac{\beta}{2}-1}^\beta (x I_\gamma) = R_{n, \gamma}^\beta x^\gamma L_{(n-1)^{(\gamma)}, \frac{\beta}{2}+1}^\beta (xI_\gamma) - \frac{n}{\beta}L_{n^{(\gamma)}, \frac{\beta}{2}-1}^\beta (x I_\gamma),
    \end{equation}
    where 
    \begin{equation*}
        R_{n, \gamma}^\beta := \frac{(-\beta)^\gamma n\beta}{(2n/\beta)_\gamma}\cdot\frac{(n\gamma)!}{(\gamma(n-1))!}.
    \end{equation*}
\end{corollary}

Finally, we obtain an explicit rational solution of the \plv V equation \eqref{eq:painleve5} from \eqref{eq:LUEpainleve}. 
\begin{corollary}\label{cor:laguerreplv}
    The solution of the \plv V equation \eqref{eq:painleve5} with the asymptotic condition \eqref{eq:painleve5cond} has the following explicit solution for $\gamma\in\mathbb{Z}$
    \begin{equation*}
        \sigma(x) = 2x R_{n, \gamma}^2 \frac{L_{(n-1)^{(\gamma)}, 2}^2(-2xI_\gamma)}{L_{n^{(\gamma)}, 0}^2(-2xI_\gamma)},
    \end{equation*}
    where $R_{n,\gamma}^2$ is the constant defined in Corollary~\ref{cor:laguerrediff}.
\end{corollary}

\section{Jacobi ensembles}\label{sec:jacobimainresult}

We define the dual parameters of $\gamma_1, \gamma_2$ similarly to those in the previous section, 
\begin{equation*}
    \tilde{\gamma_1} := \frac{2}{\beta}(\gamma_1+1) - 1, \hspace{0.5cm}\tilde{\gamma_2} := \frac{2}{\beta}(\gamma_2+1) - 1. 
\end{equation*}
In this section we use Kaneko's integral formula \cite{kaneko1993selberg} directly. Expressed in terms of $P_{\kappa, \gamma_1, \gamma_2}^\beta$, it takes the following form.
\begin{lemma}[Kaneko's integral formula]
    For $n \in \mathbb{N}$, $ \gamma_1, \gamma_2>-1$, 
    \begin{align*}
        \int_{[0, 1]^\nu}\prod_{j=1}^\nu\prod_{i=1}^n (x_i - y_j) w_{n, \gamma_1, \gamma_2}^{J, \beta} dx_1\cdots dx_n = \frac{1}{\mathcal{Z}_{n, \gamma_1+\nu, \gamma_2}^{J, \beta}}P_{n^{(\nu)}, \tilde{\gamma_1}, \tilde{\gamma_2}}^{\tilde{\beta}}(y_1, \dots, y_\nu).
    \end{align*}
\end{lemma}

An analogue of Proposition~\ref{prop:integrationconsq} holds for multivariate Jacobi polynomials. In this case, $\gamma_1$ can be altered by absorbing factors of $\prod_i x_i$ into $\prod_j\prod_i(x_i-y_j)$ with $y_j=0$, while $\gamma_2$ can be modified by incorporating factors of $\prod_i(1-x_i)$ into $\prod_j\prod_i(x_i-y_j)$ with $y_j=1$
\begin{proposition}
    For any $n, \nu\in\mathbb{N}$, $\beta>0, \gamma_1, \gamma_2>-1$, $a, b\in\mathbb{N}$, we have
    \begin{equation*}
        P_{n^{(\nu)}, \gamma_1, \gamma_2}^\beta(x_1, \dots, x_\nu)\propto P_{n^{(\nu+a+b)}, \gamma_1-\frac{\beta a}{2}, \gamma_2-\frac{\beta b}{2}}^\beta(x_1, \dots, x_\nu, \underbrace{0, \dots, 0}_a, \underbrace{1, \dots, 1}_b),
    \end{equation*}
    where $\gamma_1 - \frac{\beta a}{2}, \gamma_2 - \frac{\beta b}{2} > -1$. The ratio is determined by evaluating both polynomials at zero.
\end{proposition}

As in Section~\ref{sec:Laguerremainresult}, we study the distribution and density of the smallest eigenvalue of the J$\beta$E. For the cumulative distribution $F_{n, \gamma_1, \gamma_2}^{J, \beta}$, we obtain
\begin{align*}
    1 - F_{n, \gamma_1, \gamma_2}^{J, \beta}(x) &= \mathcal{Z}_{n, \gamma_1, \gamma_2}^{J, \beta} \int_{[x, 1]^n}\prod_{1\leq i < j \leq n}|x_i -x_j|^\beta\prod_{i=1}^n x_i^{\gamma_1} (1-x_i)^{\gamma_2} dx_1\cdots dx_n \\
    &= \mathcal{Z}_{n, \gamma_1, \gamma_2}^{J, \beta}(1-x)^{n(1+\gamma_1+\gamma_2+\beta(n-1)/2)}\\
    &\hspace{1cm}\times\int_{[0, 1]^n}\prod_{i<j}|t_i - t_j|^\beta \prod_{i=1}^n (1-t_i)^\gamma_2 \prod_{i=1}^n\left(t_i + \frac{x}{1-x}\right)^{\gamma_1}dt_1\cdots dt_n\\
    &= (1-x)^{n(1+\gamma_1+\gamma_2+\beta(n-1)/2)} P_{n^{(\gamma_1)}, \frac{2}{\beta}-1, \tilde{\gamma_2}}^{\tilde{\beta}}\left(-\frac{x}{1-x}I_{\gamma_1}\right),
\end{align*}
where we used the change of variables $t_i = (x_i - x)/(1-x)$ for all $i$. For the marginal density function $f_{n, \gamma_1, \gamma_2}^{J, \beta}$, we have
\begin{align*}
    f_{n, \gamma_1, \gamma_2}^{J, \beta}(x) &= n\int_{[x, 1]^{n-1}}f_{n, \gamma_1, \gamma_2}^{J, \beta}(x_1, \dots, x_{n-1}, x) dx_1\cdots dx_{n-1}\\
    &= n \mathcal{Z}_{n, \gamma_1, \gamma_2}^{J, \beta}x^{\gamma_1}(1-x)^{\gamma_2}\int_{[x, 1]^{n-1}} \prod_{i<j}^{n-1}|x_i - x_j|^\beta\prod_{i=1}^{n-1}x_i^{\gamma_1} (1-x_i)^{\gamma_2} |x_i - x|^\beta dx_1\cdots dx_{n-1}\\
    &= n\mathcal{Z}_{n, \gamma_1, \gamma_2}^{J, \beta}x^{\gamma_1}(1-x)^{n(1+\gamma_1+\gamma_2+\beta (n-1)/2)-(\gamma_1+1)}\\
    &\hspace{1cm}\times \int_{[0, 1]^{n-1}}\prod_{i<j}^{n-1}|t_i - t_j|^\beta \prod_{i=1}^{n-1}\left(t_i + \frac{x}{1-x}\right)^{\gamma_1} (1-t_i)^{\gamma_2} t_i^\beta dt_1\cdots dt_{n-1},
\end{align*}
where we used the same change of variables. The integral in the last expression becomes
\begin{align*}
    \int_{[0, 1]^{n-1}}&\prod_{i<j}^{n-1}|t_i - t_j|^\beta \prod_{i=1}^{n-1}\left(t_i + \frac{x}{1-x}\right)^{\gamma_1} (1-t_i)^{\gamma_2} t_i^\beta dt_1\cdots dt_{n-1} \\
    &= \int_{[0, 1]^{n-1}}w_{n-1, \beta, \gamma_2}^{J, \beta}(t_1, \dots, t_{n-1})\prod_{i=1}^{n-1}\left(t_i + \frac{x}{1-x}\right)^{\gamma_1}dt_1\cdots dt_{n-1}\\
    &= \frac{1}{\mathcal{Z}_{n-1, \gamma_1+\beta, \gamma_2}^{J, \beta}}P_{(n-1)^{(\gamma_1)}, \frac{2}{\beta}+1, \tilde{\gamma_2}}^{\tilde{\beta}}\left(-\frac{x}{1-x} I_{\gamma_1}\right),
\end{align*}
by Kaneko's formula. Thus, we obtain the following theorem for the J$\beta$E smallest eigenvalue. 
\begin{theorem}\label{thm:jacobimainresult}
    Let $\beta>0$ and $\gamma_1, \gamma_2>-1$, $\gamma_1\in\mathbb{Z}$. The smallest eigenvalue distribution $F_{n, \gamma_1, \gamma_2}^{J, \beta}$ and density $f_{n, \gamma_1, \gamma_2}^{J, \beta}$ of the $n\times n$ J$\beta$E with parameters $\gamma_1, \gamma_2$ can be written in terms of the multivariate Jacobi polynomial indexed by square partitions
    \begin{gather}
        F_{n, \gamma_1, \gamma_2}^{J, \beta}(x) = 1 - (1-x)^{n(1+\gamma_1+\gamma_2+\beta(n-1)/2)} P_{n^{(\gamma_1)}, \frac{2}{\beta}-1, \tilde{\gamma_2}}^{\tilde{\beta}}\left(-\frac{x}{1-x}I_{\gamma_1}\right),\\
        f_{n, \gamma_1, \gamma_2}^{J, \beta}(x) = \frac{n\mathcal{Z}_{n, \gamma_1, \gamma_2}^{J, \beta}}{\mathcal{Z}_{n-1, \gamma_1+\beta, \gamma_2}^{J, \beta}}x^{\gamma_1}(1-x)^{n(1+\gamma_1+\gamma_2+\beta (n-1)/2)-(\gamma_1+1)}\\ \nonumber
        \hspace{7cm}\times P_{(n-1)^{(\gamma_1)}, \frac{2}{\beta}+1, \tilde{\gamma_2}}^{\tilde{\beta}}\left(-\frac{x}{1-x} I_{\gamma_1}\right).
    \end{gather}
\end{theorem}
Once more, using the elementary relationship $\frac{d}{dx}F(x) = f(x)$ and the change of variable $x\mapsto -x/(1-x)$, we derive the following differentiation formula
\begin{align*}
    n(1+\gamma_1+\gamma_2+\frac{\beta(n-1)}{2})P_{n^{(\gamma_1)}, \frac{2}{\beta}-1, \tilde{\gamma_2}}^{\tilde{\beta}}(x) &+ (1-x)\frac{d}{dx}P_{n^{(\gamma_1)}, \frac{2}{\beta}-1, \tilde{\gamma_2}}^{\tilde{\beta}}(x)\\
    &= \frac{(-1)^{\gamma_1}n\mathcal{Z}_{n, \gamma_1, \gamma_2}^{J, \beta}}{\mathcal{Z}_{n-1, \gamma_1+\beta, \gamma_2}^{J, \beta}} x^{\gamma_1} P_{(n-1)^{(\gamma_1)}, \frac{2}{\beta}+1, \tilde{\gamma_2}}^{\tilde{\beta}}(x).
\end{align*}
Reorganizing some of the parameters ($\tilde\beta\mapsto \beta, \tilde{\gamma_2}\mapsto\gamma_2$) we obtain the following differentiation formula for multivariate Jacobi polynomials indexed by square partitions.
\begin{corollary}\label{cor:jacobidiff}
    For parameters $\beta>0$, $\gamma_1$ we have
    \begin{align}\nonumber
        (1-x)&\frac{d}{dx}P_{n^{(\gamma_1)}, \frac{\beta}{2}-1, \gamma_2}^\beta (x) \\
        &= \frac{(-1)^{\gamma_1}n\mathcal{Z}_{n, \gamma_1, \tilde{\gamma_2}}^{J, \tilde{\beta}}}{\mathcal{Z}_{n-1, \gamma_1+\tilde{\beta}, \tilde{\gamma_2}}^{J, \tilde{\beta}}} x^{\gamma_1} P_{(n-1)^{(\gamma_1)}, \frac{\beta}{2}-1, \gamma_2}^\beta(x) - n(1+\gamma_1+\tilde{\gamma_2}+\frac{2(n-1)}{\beta})P_{n^{(\gamma_1)}, \frac{\beta}{2}-1, \gamma_2}^\beta (x),
    \end{align}
    where the constant term simplifies to
    \begin{equation*}
         \frac{(-1)^{\gamma_1}n\mathcal{Z}_{n, \gamma_1, \tilde{\gamma_2}}^{J, \tilde{\beta}}}{\mathcal{Z}_{n-1, \gamma_1+\tilde{\beta}, \tilde{\gamma_2}}^{J, \tilde{\beta}}} = \frac{\Gamma(1+\frac{2}{\beta})\Gamma(1+\gamma_1+\frac{2}{\beta}(\gamma_2+n))\Gamma(1+\gamma_1+\frac{2}{\beta}n)}{\Gamma(1+\frac{2}{\beta}n)\Gamma(1+\gamma_1)\Gamma(1+\gamma_1+\frac{2}{\beta})\Gamma(\frac{2}{\beta}(\gamma_2+n))}. 
    \end{equation*}
\end{corollary}

We finally consider the \plv VI differential equation \cite[Eq. (4.26)]{haine1999jacobi} associated with the Jacobi unitary ensemble. After rescaling the variables, we have the following result.
\begin{corollary}\label{cor:jacobiplv}
    The \plv VI equation \cite[Eq.~(4.26)]{haine1999jacobi} has the following explicit rational function solution for $\gamma_1\in\mathbb{Z}$
    \begin{equation*}
        \sigma(x) = \frac{1}{2}g(x) - \frac{(2n+\gamma_1+\gamma_2)^2}{4}x + \frac{2n^2 + 2(\gamma_1+\gamma_2)n + \gamma_1(\gamma_1+\gamma_2)}{4},
    \end{equation*}
    where $g(x)$ is defined in terms of the ratio of two multivariate Jacobi polynomials
    \begin{equation*}
        g(x) = \frac{2\Gamma(1+\gamma_1+n)\Gamma(1+n+\gamma_1+\gamma_2)}{\Gamma(n)\Gamma(1+\gamma_1)\Gamma(\gamma_1+2)\Gamma(\gamma_2+n)}\cdot \frac{x^{\gamma_1+1}}{(1-x)^{\gamma_1}}\cdot \frac{P_{(n-1)^{(\gamma_1)}, 2, \gamma_2}^2\left(\frac{-x}{1-x}I_{\gamma_1}\right)}{P_{n^{(\gamma_1)}, 0, \gamma_2}^2\left(\frac{-x}{1-x}I_{\gamma_1}\right)}. 
    \end{equation*}
\end{corollary}

\section{Numerical experiments}\label{sec:numerics}

In this section we perform numerical experiments and verify Theorems~\ref{thm:laguerremainresult} and~\ref{thm:jacobimainresult}. In the following experiments, we sample $10^6$ smallest eigenvalues of the L$\beta$E or J$\beta$E and compare the statistics of the samples with the formulas obtained in Theorems~\ref{thm:laguerremainresult} and~\ref{thm:jacobimainresult}. Multivariate Laguerre and Jacobi polynomials are computed symbolically using the algorithm proposed in \cite{dumitriu2007mops}, with slight modifications to match our normalizations in \eqref{eq:lbedensity} and \eqref{eq:jbedensity}. Once again the explicit expressionso of the polynomials used in this section are listed in Appendix~\ref{sec:examples}. 

\subsection{Laguerre ensemble}

We obtain samples of L$\beta$E for $\beta>0$ with the bidiagonal matrix model proposed in \cite{dumitriu2002matrix}. 

\subsubsection{Parameters $\beta=\frac{5}{2}$, $n=4$, $\gamma=2$}

\begin{figure}[h]
    \includegraphics[width=0.49\textwidth]{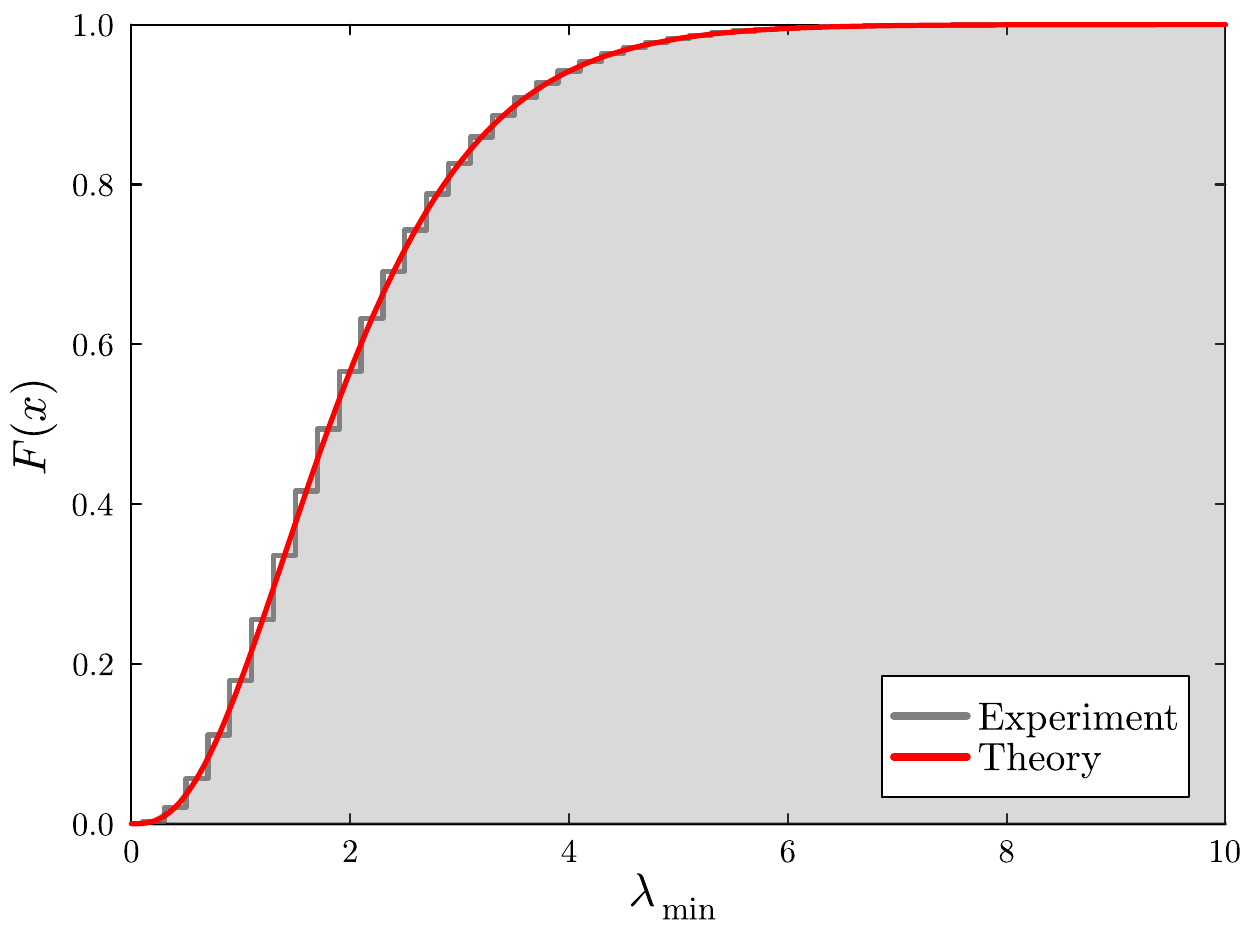}
    \includegraphics[width=0.49\textwidth]{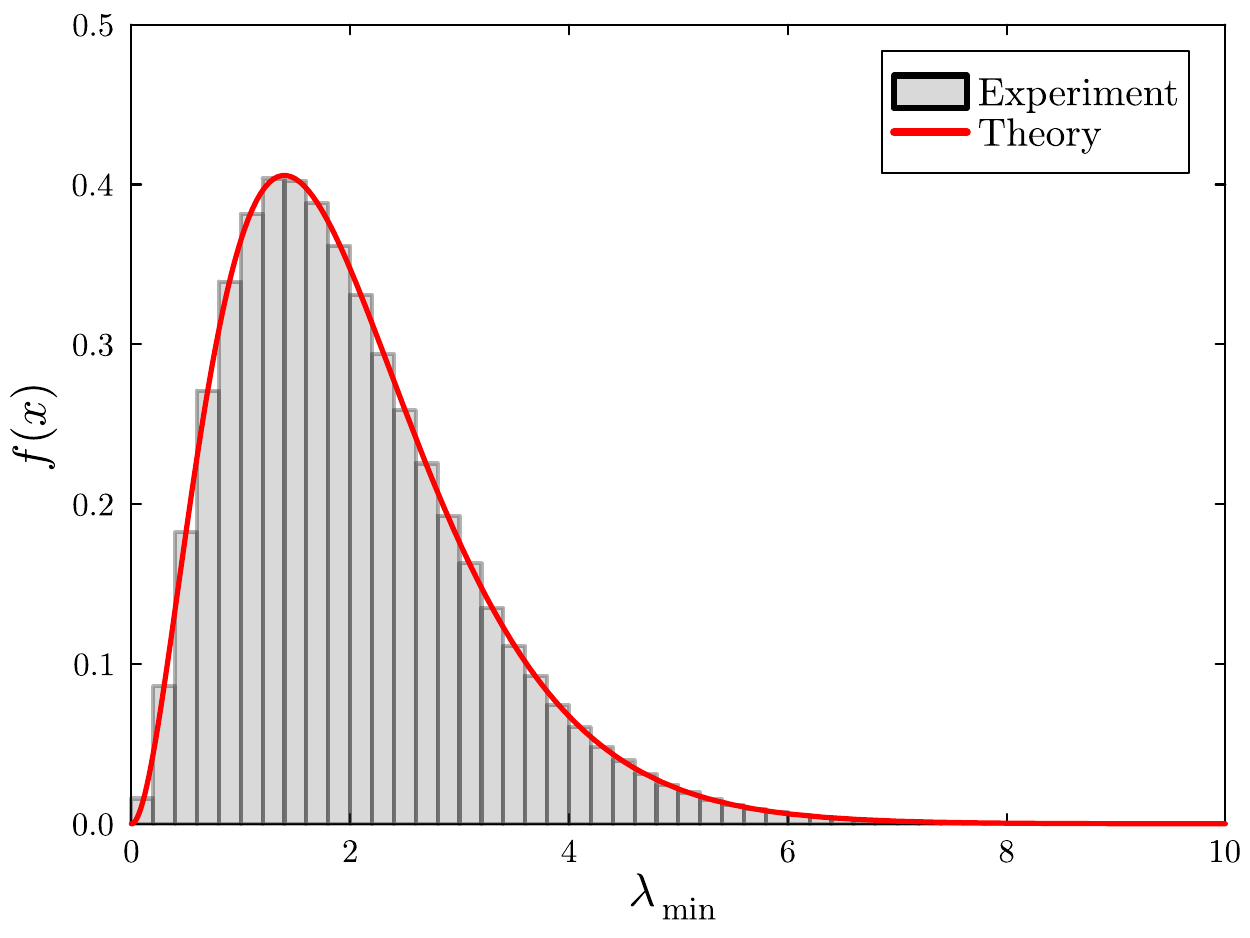}
    \caption{The empirical CDF (left) and PDF (right) of the smallest eigenvalue of the L$\beta$E, $\beta=\frac{5}{2}$, $n=4$, $\gamma=2$ are given in gray. The red curves represents Theorem~\ref{thm:laguerremainresult}.}
    \label{fig:firstcomparison}
\end{figure}

By Theorem~\ref{thm:laguerremainresult}, we need to compute 
\begin{equation*}
    L_{[4, 4], -\frac{1}{5}}^{\frac{4}{5}}\left(-\frac{4x}{5}I_2\right), \hspace{1cm}L_{[3, 3], \frac{9}{5}}^{\frac{4}{5}}\left(-\frac{4x}{5}I_2\right).
\end{equation*}
They are written in \eqref{eq:cdf1experiment} and \eqref{eq:pdf1experiment}. See Figure~\ref{fig:firstcomparison} for the comparison. 

\subsubsection{Parameters $\beta=e$, $n=3$, $\gamma=2$}
\begin{figure}[h]
    \includegraphics[width=0.49\textwidth]{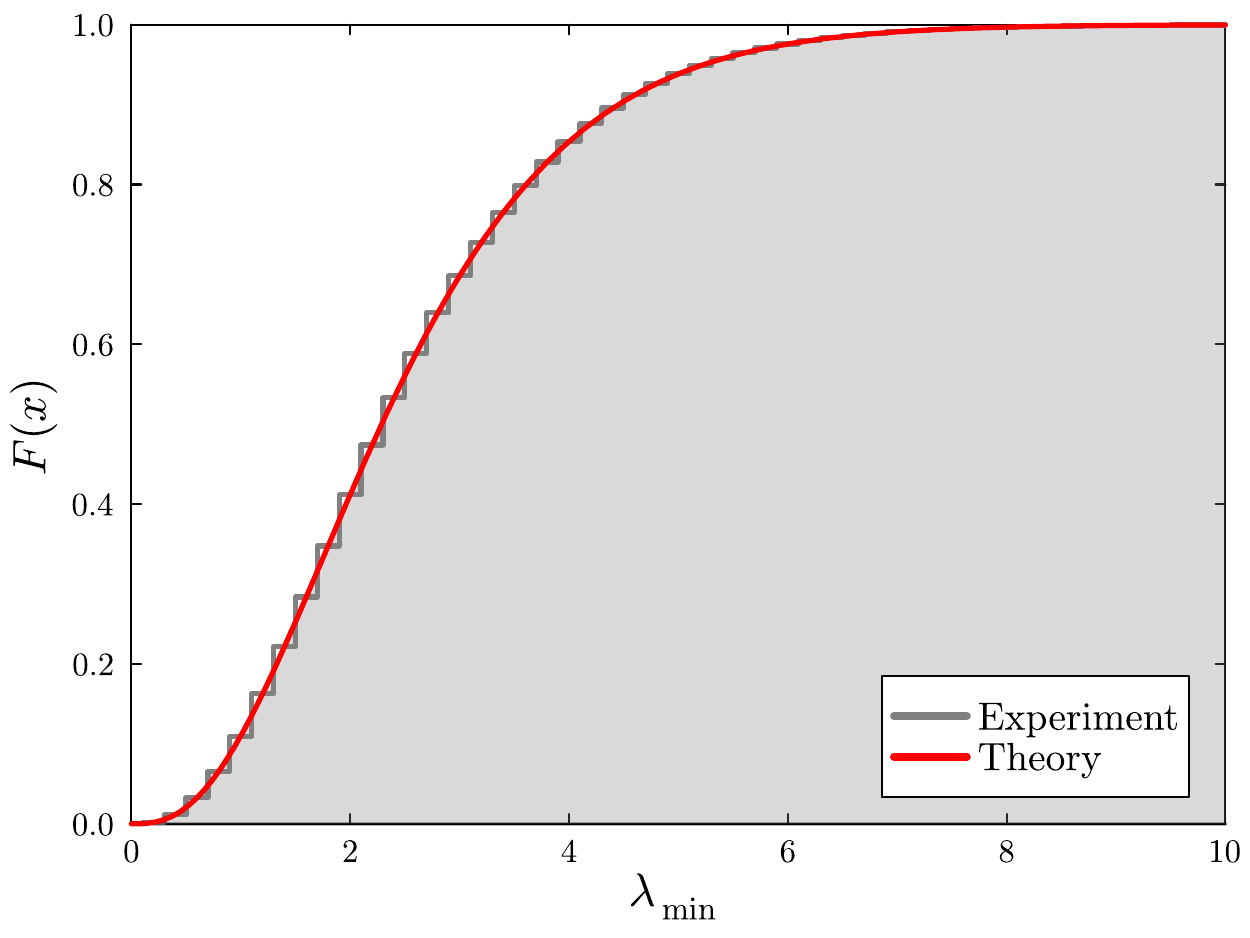}
    \includegraphics[width=0.49\textwidth]{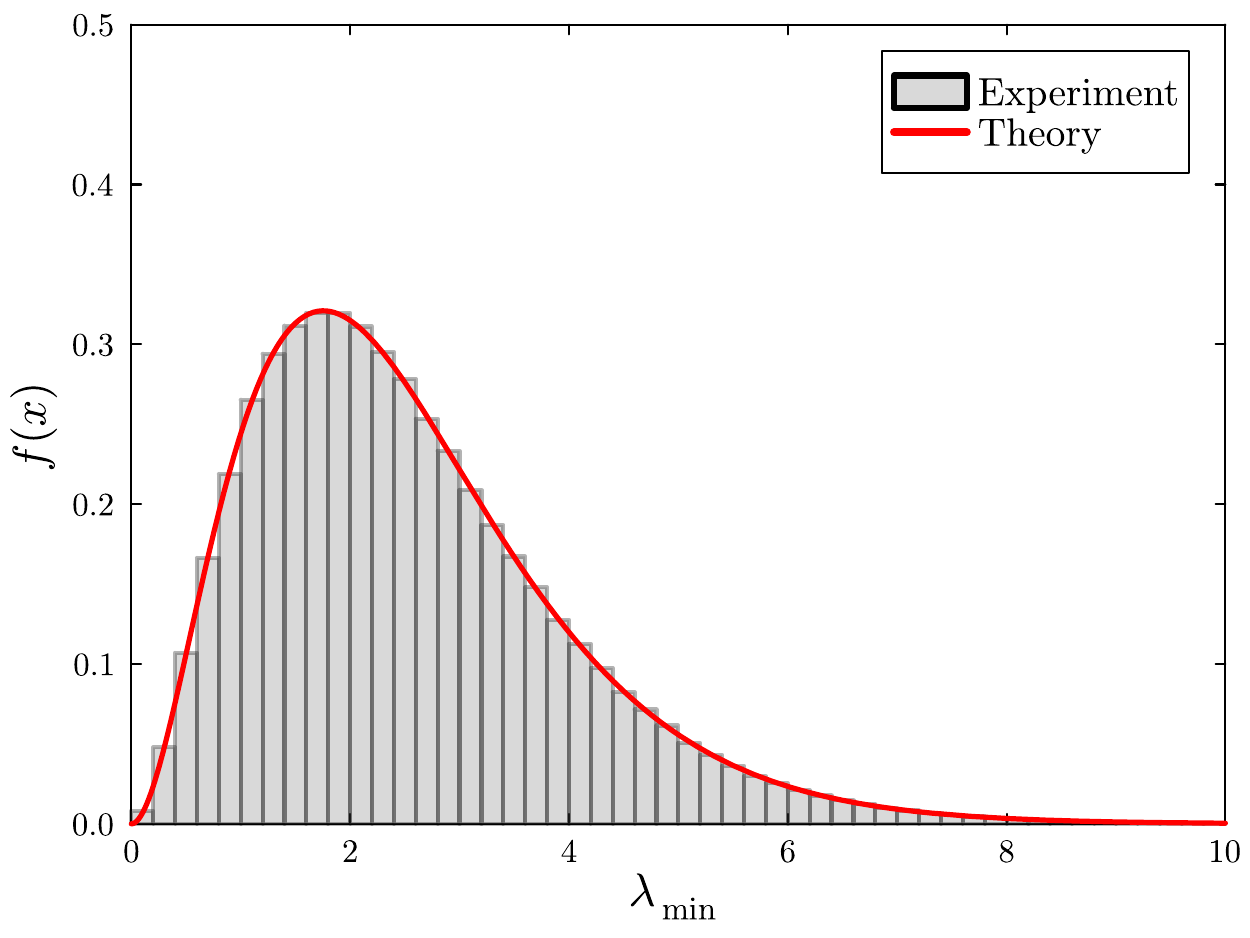}
    \caption{The empirical CDF (left) and PDF (right) of the smallest eigenvalue of the L$\beta$E, $\beta=e$, $n=3$, $\gamma=2$ are given in gray. The red curves represents Theorem~\ref{thm:laguerremainresult}..}
    \label{fig:secondcomparison}
\end{figure}

Note that multivariate Laguerre polynomials can be computed with symbolic parameters from the algorithm in \cite{dumitriu2007mops}. We use two multivariate Laguerre polynomials, 
\begin{equation*}
    L_{[3,3], \frac{2}{e}-1}^{\frac{4}{e}}(-\frac{2x}{e}I_2), \hspace{1cm}L_{[2, 2], \frac{2}{e}+1}^{\frac{4}{e}}(-\frac{2x}{e}I_2),
\end{equation*}
which are given in \eqref{eq:cdf2experiment} and \eqref{eq:pdf2experiment}. See Figure~\ref{fig:secondcomparison} for the comparison. The density function $f_{n, \gamma}^{L, \beta}$ with this set of parameters is also reported in \cite[Table 1]{kumar2019recursion}, which agrees with our result. 

\subsection{Jacobi ensemble}
We obtain samples of the J$\beta$E with the tridiagonal matrix model proposed in \cite{killip2004matrix}. 

\subsubsection{Parameters $\beta=4/3$, $n=3$, $\gamma_1 = 2$, $\gamma_2 = 2/3$}
\begin{figure}[h]
    \includegraphics[width=0.49\textwidth]{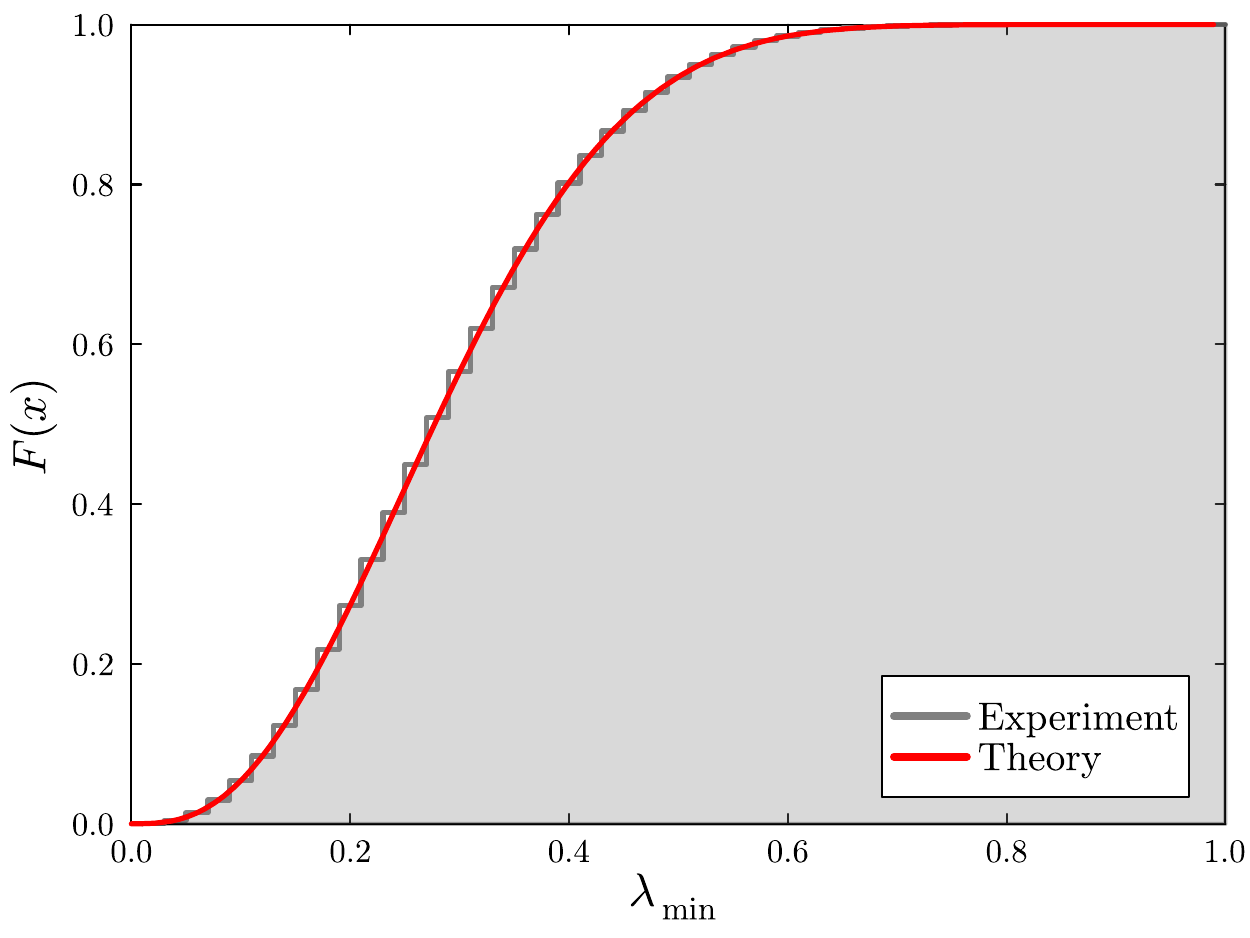}
    \includegraphics[width=0.49\textwidth]{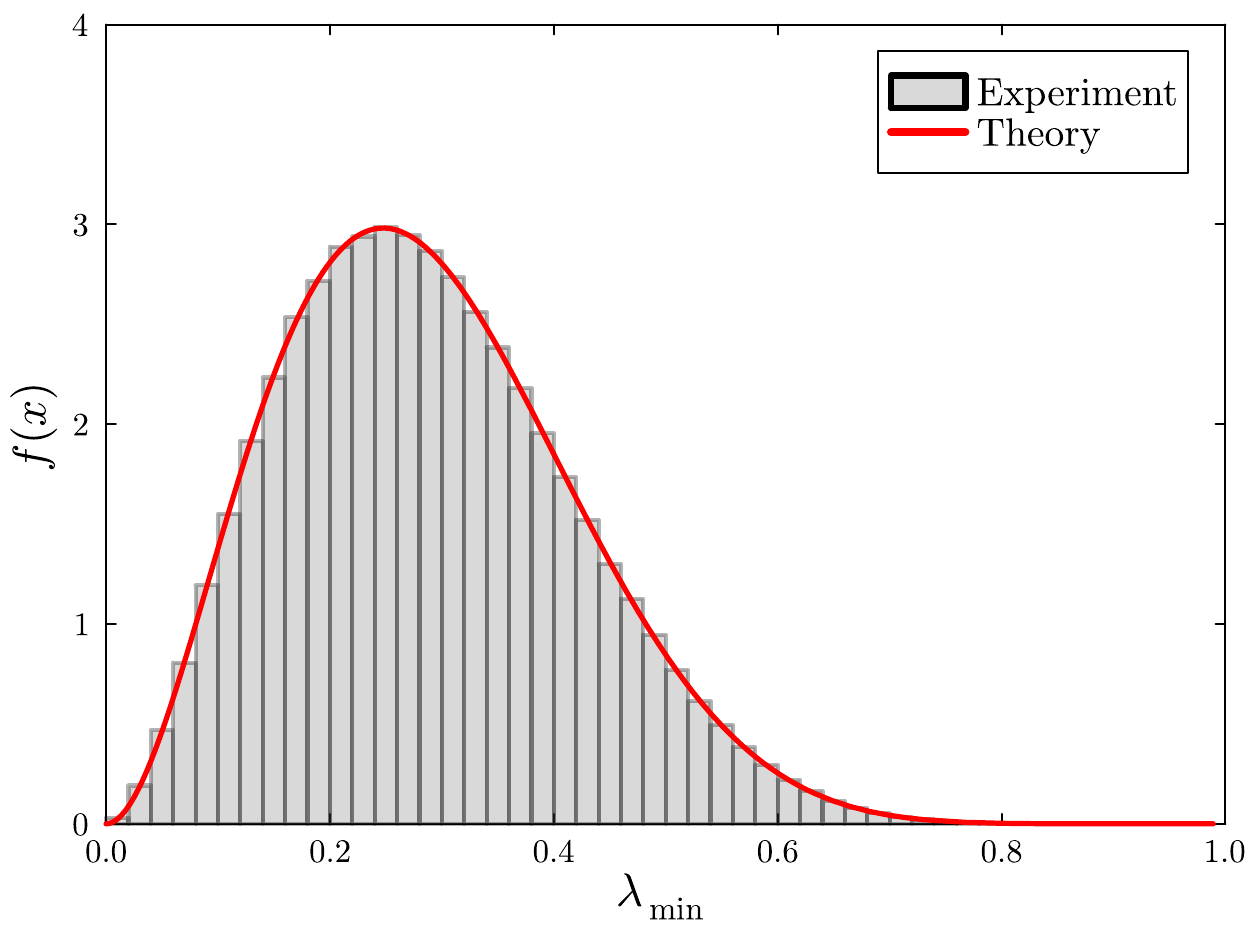}
    \caption{The empirical CDF (left) and PDF (right) of the smallest eigenvalue of the J$\beta$E, $\beta=4/3$, $n=3$, $\gamma_1 = 2$, $\gamma_2 = 2/3$ are given in gray. The red curves represents Theorem~\ref{thm:jacobimainresult}.}
    \label{fig:thirdcomparison}
\end{figure}

By Theorem~\ref{thm:jacobimainresult} we need to compute
\begin{equation*}
    P_{[3,3], \frac{1}{2}, \frac{3}{2}}^3\left(xI_2\right), \hspace{1cm}P_{[2,2], \frac{5}{2}, \frac{3}{2}}^3\left(xI_2\right),
\end{equation*}
which are given in \eqref{eq:cdf3experiment} and \eqref{eq:pdf3experiment}. See Figure~\ref{fig:thirdcomparison} for the comparison.

\subsubsection{Parameters $\beta=1/2$, $n = 2$, $\gamma_1=3$, $\gamma_2=1/4$}
\begin{figure}[h]
    \includegraphics[width=0.49\textwidth]{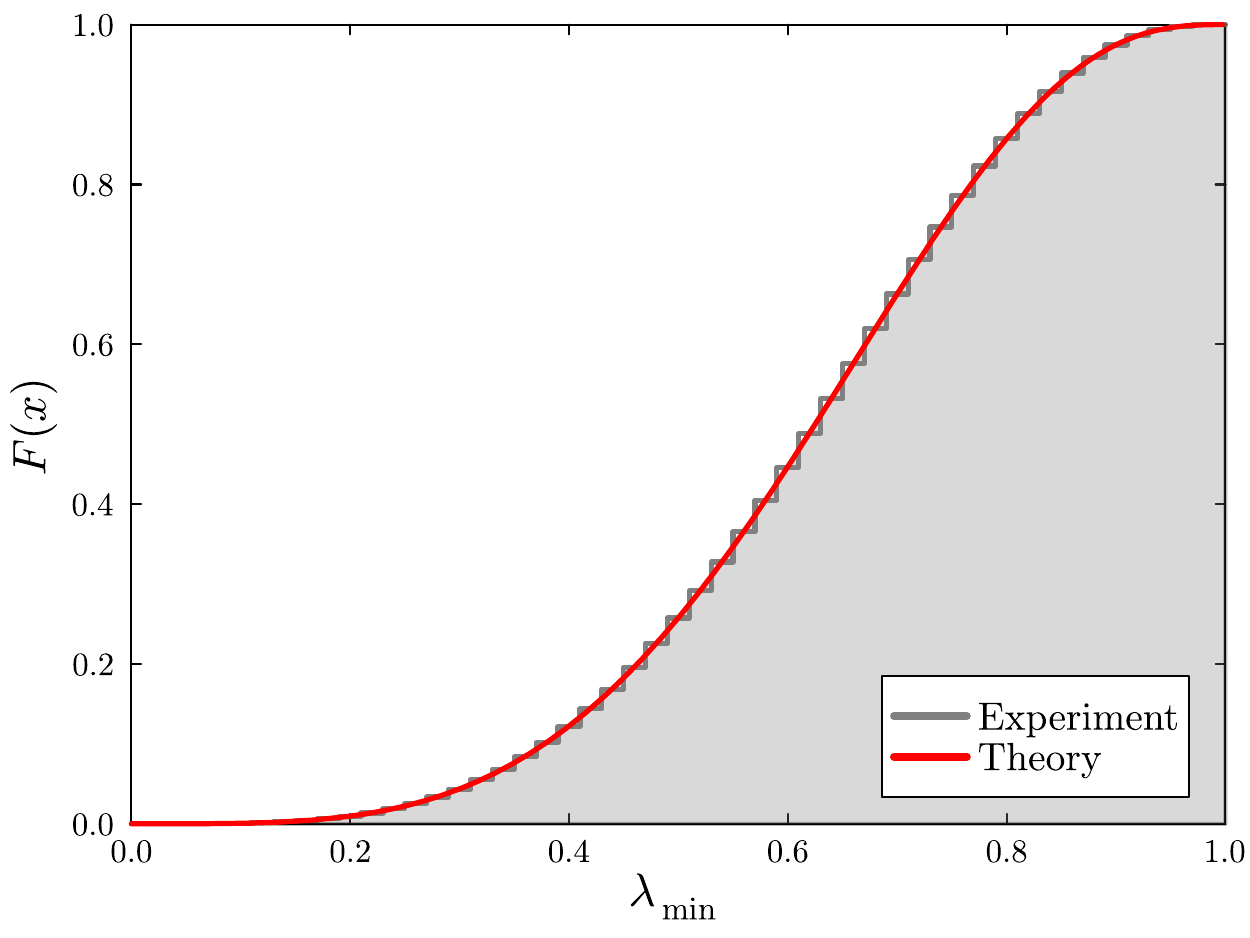}
    \includegraphics[width=0.49\textwidth]{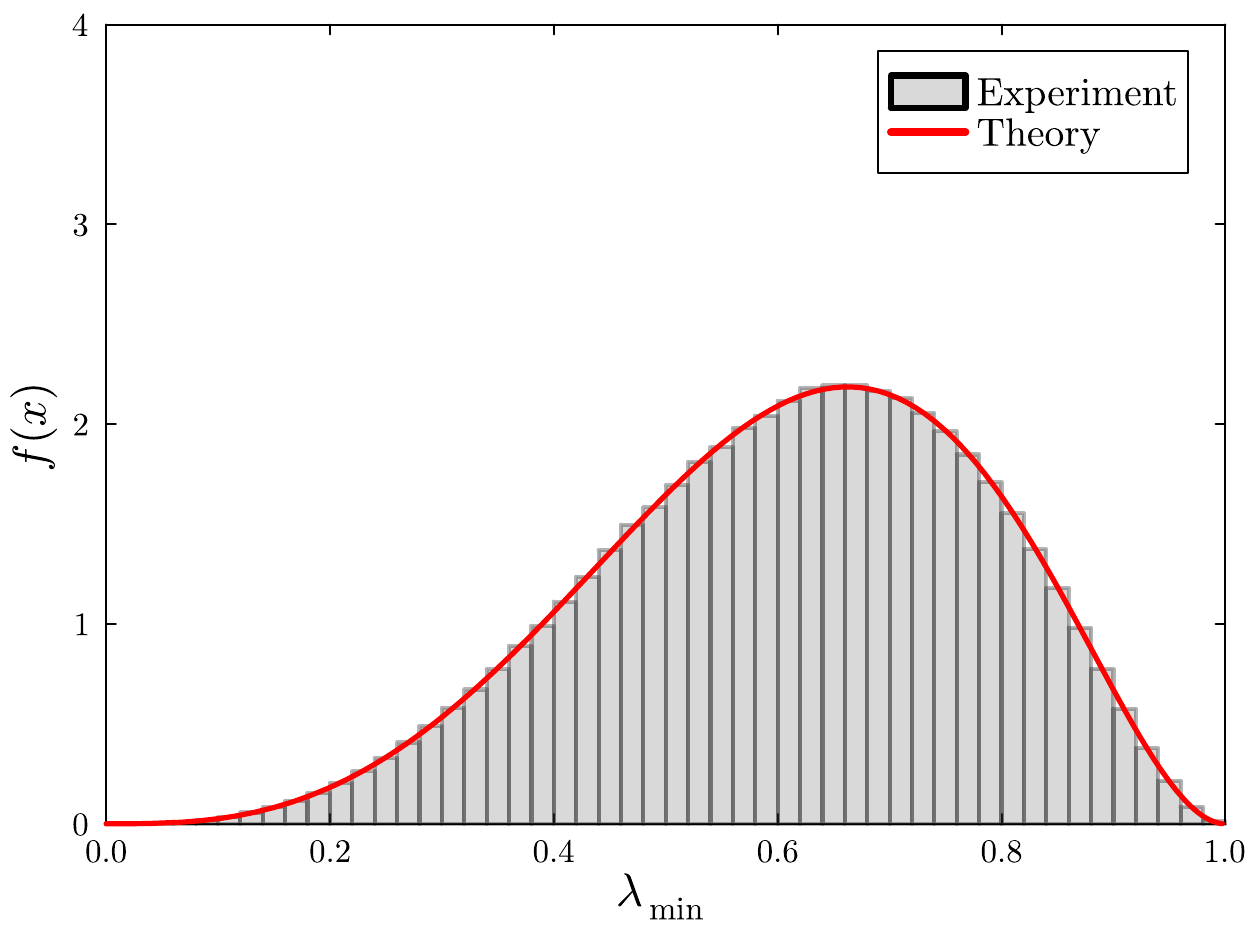}
    \caption{The empirical CDF (left) and PDF (right) of the smallest eigenvalue of the J$\beta$E, $\beta=1/2$, $n = 2$, $\gamma_1=3$, $\gamma_2=1/4$ are given in gray. The red curves represents Theorem~\ref{thm:jacobimainresult}.}
    \label{fig:fourthcomparison}
\end{figure}

By Theorem~\ref{thm:jacobimainresult} we need to compute
\begin{equation*}
    P_{[2,2,2], 3, 4}^8\left(xI_3\right), \hspace{1cm}P_{[1,1,1], 5, 4}^8\left(xI_3\right),
\end{equation*}
which are given in \eqref{eq:cdf4experiment} and \eqref{eq:pdf4experiment}. See Figure~\ref{fig:fourthcomparison} for the comparison.

\subsection*{Acknowledgements}
The author greatly appreciates Alan Edelman for many helpful suggestions and insights. 

\bibliographystyle{alpha}
\bibliography{bibliography}

\appendix
\section{Explicit multivariate polynomials}\label{sec:examples}

We begin by presenting a few examples of multivariate Laguerre and Jacobi polynomials, which serve as reference points for our normalization.
\begin{itemize}
    \item Laguerre, $\beta=3$, $\gamma=1$, $\kappa=[2,1]$
    \begin{align*}
        L_{[2, 1], 1}^3 (x, y, z) &= -\frac{3}{14}\big(x^2y + x^2z + xy^2 + xz^2 + y^2z + yz^2\big) -\frac{27}{56}\,xyz \\
        & + 3\,(x^2+y^2+z^2) + \frac{27}{2}\,(xy+xz+yz) - \frac{297}{2}\,(x+y+z) + 1485.
    \end{align*}
    \item Laguerre, $\beta=\frac{5}{2}$, $\gamma=\frac{3}{2}$, $\kappa=[1,1]$
    \begin{align*}
        L_{[3,2],\frac{3}{2}}^{\frac{5}{2}}(x,y)=\frac{15}{442}&\big(-4(x^3y^2+x^2y^3)+56(x^3y+xy^3)+236x^2y^2-140(x^3+y^3)\\
        &-2688(x^2y+xy^2)+30268xy+6440(x^2+y^2)-76475(x+y)+229425\big)
    \end{align*}
    \item Jacobi, $\beta=3$, $\gamma_1=\frac{3}{2}$, $\gamma_2=\frac{1}{2}$, $\kappa=[2,1]$
    \begin{align*}
        P_{[2,1],\frac{3}{2},\frac{1}{2}}^3(x,y) = -\frac{117}{50}(x^2y + xy^2) + \frac{117}{80}(x^2 + y^2) + \frac{1083}{600}xy-  \frac{5}{2}(x + y) + 1
    \end{align*}
\end{itemize}

\subsection*{Polynomials used in previous sections} 

Following are the explicit expressions of the polynomials that are used in previous sections. 
\begin{align} 
    L_{[3,3,3],5}^8&(-4x I_3) = \frac{1}{55}( 8192 x^{9} + 294912 x^{8} + 5160960 x^{7} + 56082432 x^{6} + 411844608 x^{5} \nonumber \\
    &+ 2124251136 x^{4} + 7709440512 x^{3} + 19211157408 x^{2} + 30656102400 x + 23843635200) \label{eq:kumarcomparison}
\end{align}

\begin{align}
L_{[4,4,4],3}^8(-4x I_3) = &1024(4x^{12} + 168x^{11} + 3444x^{10} + 44604x^9 + 401814x^8 \nonumber\\ \label{eq:laguerresectionexF}
&+ 2644488x^7 + 13020084x^6 + 48440700x^5 + 136070550x^4 \\ 
&+ 283783500x^3 + 425675250x^2 + 425675250x + 212837625)    \nonumber
\end{align}

\begin{align} 
    L_{[4,4], -\frac{1}{5}}^{\frac{4}{5}}\left(-\frac{4x}{5} I_2\right) &= \frac{1}{21375}(256x^{8} + 19440x^{7} + 592896x^{6} + 921600x^{5}\nonumber \\ \label{eq:cdf1experiment}
        &+ 7759872x^{4} + 35203200x^{3} + 823260480x^{2} + 823260480x + 410299200)
\end{align}

\begin{align}
    L_{[3,3], \frac{9}{5}}^{\frac{4}{5}}\left(-\frac{4x}{5} I_2\right) = \frac{256}{9975}(x^{6} + 72x^{5} &+ 2048x^{4} + 28920x^{3}\nonumber\\ \label{eq:pdf1experiment}
    &+ 213060x^{2} + 773550x + 1159500)
\end{align}

\begin{align} 
   L_{[3,3], \frac{2}{e}-1}^{\frac{4}{e}}&\left(-\frac{2x}{e}I_2\right) = \frac{60}{e^3(e+1)(e+2)(3e+2)} \nonumber \\ \label{eq:cdf2experiment}
   & \times \Big(x^6 + 12(e+1)x^5 + 24(e+1)(2e+3)x^4 + 8(e+1)(9e+16)(e+2)x^3 \\ \nonumber
   & + 36(e+1)(e+2)^2(e+4)x^2 + 48(e+1)(e+2)^2(e+4)x + 32(e+1)(e+2)^2(e+4)\Big)
\end{align}

\begin{align}
    L_{[2, 2], \frac{2}{e}+1}^{\frac{4}{e}}\left(-\frac{2x}{e}I_2\right) &= \frac{6}{e^2(e+1)(e+2)}\Big(x^4 + 4(3e+2)x^3 + 16(e+1)(3e+2)x^2 \nonumber\\ \label{eq:pdf2experiment}
    &\hspace{1cm}+ 8(e+1)(3e+2)(3e+4)x + 4(e+1)(e+2)(3e+2)(3e+4)\Big)
\end{align}

\begin{align}\label{eq:cdf3experiment}
    P_{[3,3], \frac{1}{2}, \frac{3}{2}}^3(x I_2) = \frac{1292}{5}x^6-\frac{6783}{10}x^5+\frac{7089}{10}x^4-374x^3+105x^2-15x+1
\end{align}

\begin{align}\label{eq:pdf3experiment}
    P_{[2, 2], \frac{5}{2}, \frac{3}{2}}^3(xI_2) = \frac{1}{135}\big(1292x^{4} - 2907x^{3} + 2448x^{2} - 918x + 135\big)
\end{align}

\begin{align}\label{eq:cdf4experiment}
    P_{[2,2,2], 3, 4}^8(xI_3) = \frac{1}{208}(7315x^6 - 19950x^5 + 24762x^4 - 17472x^3 + 7488x^2 - 1872x + 208)
\end{align}

\begin{align}\label{eq:pdf4experiment}
    P_{[1,1,1],5, 4}^8(x I_3) = -\frac{1}{616}(2299x^{3} - 3762x^{2} + 2508x - 616)
\end{align}

\end{document}